\documentstyle[preprint,aps]{revtex}

\begin{document}

\draft

%\preprint{xxx-yy/95}

\title{Spinning BTZ Black Hole versus Kerr Black Hole : \\
       A Closer Look}

\author{Hongsu Kim \footnote{e-mail : hongsu@sirius.kyungpook.ac.kr}}

\address{Department of Astronomy and Atmospheric Sciences \\
Kyungpook National University, Taegu 702-701, KOREA}

\date{August, 1998}

\maketitle

\begin{abstract}
By applying Newman's algorithm, the AdS$_{3}$ rotating black hole solution
is ``derived" from the nonrotating black hole solution of 
Ba$\tilde{\rm{n}}$ados, Teitelboim, and Zanelli (BTZ). 
The rotating BTZ solution derived in this fashion is given in
``Boyer-Lindquist-type'' coordinates whereas the form of the solution
originally given by BTZ is given in a kind of an ``unfamiliar'' 
coordinates which are related to each other by a transformation of time 
coordinate alone. The relative physical meaning between these two time 
coordinates is carefully studied. Since the Kerr-type and Boyer-Lindquist-
type coordinates for rotating BTZ solution are newly found via Newman's
algorithm, next, the transformation to Kerr-Schild-type coordinates is
looked for. Indeed, such transformation is found to exist. And in this
Kerr-Schild-type coordinates, truely maximal extension of its global
structure by analytically continuing to ``antigravity universe'' region
is carried out.

\end{abstract}


\narrowtext
%\twocolumn

%%%%%%%%%%%%%%%%%%%%%%%%%%%%%%%%%%%%%%%%%%%%%%%%%%%%%%%%%%%%%%%%%%%%%%%%%%%%%%%

%\newpage

%%%%%%%%%%%%%%%%%%%%%%%%%%%%%%%%%%%%%%%%%%%%%%%%%%%%%%%%%%%%%%%%%%%%%%%%%%%%%%%%%%%%%%

{\bf I. Introduction}
\\
It had long been thought that black hole solutions cannot exist in 3-dim. since
there is no local gravitational attraction and hence no mechanism to confine
large densities of matter. It was, therefore, quite a surprise when 
Ba$\tilde{\rm{n}}$ados,
Teitelboim, and Zanelli (BTZ) [1]  have recently constructed the 
Anti-de Sitter (AdS$_{3}$)
spacetime solution to the Einstein equations in 3-dim. that can be interpreted
as a black hole solution. They included the negative cosmological constant in the
3-dim. vacuum Einstein theory and then found both the rotating and nonrotating
black hole solutions. In the mean time, a curious relationship between the
nonrotating and the rotating spacetime solutions of Einstein theory in 4-dim. 
also has been long known. Newman et al. [4] discovered long ago that one can 
``derive"
Kerr solution from the Schwarzschild solution in vacuum Einstein theory and 
Kerr-Newman solution from the Reissner-Nordstr$\ddot{\rm{o}}$m solution in 
Einstein-Maxwell
theory via the ``complex coordinate transformation" scheme acting on metrics
written in terms of null tetrad of basis vectors. In the present work, we 
attempt the
same derivation but in this time in 3-dim. spacetime. Namely, we see if the 
rotating
version of BTZ black hole solution can indeed be ``derived" from its 
nonrotating
counterpart via Newman's method. And in doing so, our philosophy is that the
3-dim. situation can be thought of as, say, the $\theta = \pi/2$ - slice of the
4-dim. one (where $\theta$ denotes the polar angle). 
Interestingly enough, we do end up with the rotating version of BTZ
black hole solution but in a different coordinate system from the one 
originally
employed in BTZ's solution ansatz. And as we shall see shortly, it turns out
that the rotating BTZ solution derived here in this work is given in
``Boyer-Lindquist-type'' [5] coordinates whereas the original form of the 
solution given by BTZ is given in a kind of an ``unusual'' coordinates which 
are related to
the ``familiar'' Boyer-Lidquist-type coordinates by a transformation of time
coordinate alone. We can easily understand the reason for this result as 
follows ; much as the Kerr solution ``derived'' from the Schwarzschild
solution by Newman's complex coordinate transformation method naturally
comes in Kerr coordinates [2,4] from which one can transform to
Boyer-Lindquist coordinates [5], the rotating BTZ black hole solution
derived from its nonrotating counterpart in this manner comes in
Kerr-type coordinates again from which one can make a transformation to
Boyer-Lindquist-type coordinates as well. And then one can realize that by
performing a transformation from the Boyer-Lindquist-type time coordinate
$t$ to a ``new'' time coordinate $\tilde{t}$ following the transformation law,
$\tilde{t} = t - a \phi$ (where $a$ is proportional to the angular momentum 
per unit mass and $\phi$ denotes the azimuthal angle), 
our ``derived'' rotating BTZ black hole solution can indeed be put
in the form originally given by BTZ.
As expected, the rotating BTZ black hole solution given in Boyer-Lindquist-type
coordinates takes on the structure which resembles that of Kerr solution
more closely than it is in BTZ's rather unusual time coordinate. 
However, one can readily realize that it is the BTZ's time coordinate 
$\tilde{t}$ that is the usual Killing time coordinate, not the Boyer-Lindquist-
type one.
The careful analysis of the relative physical meaning between the the two
different choices of coordinates will be presented later on in the discussion. 
Then this successful derivation of rotating BTZ black hole solution from its
nonrotating counterpart via Newman's algorithm plus the fact that the resultant 
metric comes in Kerr-type and Boyer-Lindquist-type coordinates which are
familiar ones well-known in the study of Kerr black hole solution in 4-dim.
lead us to attempt to complete the study of rotating BTZ solution in
parallel with that of Kerr solution. Recall that in addtion to Kerr and
Boyer-Lindquist coordinates, there is one more coordinate system of central
importance, i.e., the Kerr-Schild coordinates. And in this coordinates,
one can envisage peculiar features of Kerr spacetime such as the ``ring''
structure of curvature singularity and the analytic continuation of
(Boyer-Lindquist) $r$-coordinate from positive to negative values.
Therefore, we similarly look for further transformation to Kerr-Schild-type
coordinates for the rotating BTZ metric. It turns out that such
Kerr-Schild-type coordinates indeed exists and in terms of which the
rotating BTZ black hole metric takes on precisely the same form as that of 
Kerr metric sliced along the $\theta = \pi/2$ equatorial plane.
As a result, this one-to-one correspondence between the metric
(or coordinates) of rotating BTZ spacetime and that of Kerr spacetime
allows us to envisage the structure of rotating BTZ solution in
different direction from that in which the previous study of rotating
BTZ solution has been done. For example, the representation of the
rotating BTZ black hole metric in Kerr-Schild-type coordinates allows
us to realize that the global structure of the rotating hole can be
maximally extended by analytically continuing the (Boyer-Lindquist-type)
$r$-coordinate from positive (``our universe'' region) to negative values 
(``antigravity universe''region) through $r = 0$ in
a similar manner to the case of Kerr spacetime in 4-dim. 
\\
{\bf II. Derivation of the AdS$_{3}$ rotating black hole solution via 
Newman's algorithm}
\\
As mentioned earlier, Newman et al. [4] discovered curious ``derivations" of 
stationary,
axisymmetric metric solutions from static, spherically-symmetric solutions
in 4-dim. Einstein theory. To attempt the similar work in 3-dim. spacetime, we
start with the nonrotating version of BTZ black hole solution in 3-dim. as a 
``seed" 
solution to construct its rotating counterpart. The nonrotating BTZ black hole
solution written in Schwarzschild-type coordinates $(t, r, \phi)$ is given by
\begin{eqnarray}
ds^2 = (-M + {r^2\over l^2})dt^2 - (-M + {r^2\over l^2})^{-1}dr^2 - r^2d\phi^2
\end{eqnarray}
where $l$ is related to the negative cosmological constant by $l^{-2} = 
-\Lambda$
and $M$ is an integration constant that can be identified with the ADM mass 
of the black hole. Now in order to ``derive" a rotating black hole solution 
applying the
complex coordinate transformation scheme of Newman et al., we begin by assuming
that this 3-dim. black hole geometry is a $\theta = \pi/2$ - slice of a static,
spherically-symmetric 4-dim. geometry given by
\begin{eqnarray}
ds^2_{4} = \lambda^{2}(r)dt^2 - \lambda^{-2}(r)dr^2 - r^2 (d\theta^2 + 
\sin^2 \theta d\phi^2)
\end{eqnarray} 
with $\lambda^{2}(r) = (-M+r^{2}/l^{2})$. The essential reason for this 
``dimensional continuation" is to introduce the null tetrad system of vectors
on which Newman's complex coordinate transformation method is technically 
based.
We do not, however, ask nor demand that this 4-dim. geometry be an explicit
solution of Einstein equation in 4-dim. as well. We just demand that only its
$\theta = \pi/2$ - slice be a solution of Einstein equation in 3-dim.
Remarkably, then, upon the series of operations ; dimensional continuation 
$\rightarrow $ Newman's derivation method $\rightarrow $ dimensional reduction
by setting $\theta = \pi/2$, we end up with a legitimate rotating black hole
solution to 3-dim. Einstein equation as we shall see shortly.
\\
Consider now the transformation to the Eddington-Finkelstein-type retarded 
null coordinates $(u, r, \phi)$ defined by $u = t - r_{\ast}$, with 
$r_{\ast} = \int dr (g_{rr}/-g_{tt})^{1/2}$. In terms of these null 
coordinates, the nonrotating BTZ black hole metric takes the form
\begin{eqnarray}
ds^2_{4} = \lambda^{2}(r)du^2 + 2du dr - r^2 (d\theta^2 +
\sin^2 \theta d\phi^2).
\end{eqnarray}
Into this 4-dim. Riemannian space we next introduce a tetrad system of vectors 
$l_{\mu}$,
$n_{\mu}$, $m_{\mu}$ and $\bar{m}_{\mu}$ (with $\bar{m}_{\mu}$ being the 
complex 
conjugate of $m_{\mu}$) satisfying the following orthogonality property
$l_{\mu}n^{\mu} = - m_{\mu}\bar{m}^{\mu} = 1$ with all other scalar products
vanishing. In terms of this null tetrad of basis vectors, the spacetime 
metric is written as $g^{\mu \nu} = l^{\mu}n^{\nu} + n^{\mu}l^{\nu} - 
m^{\mu}\bar{m}^{\nu} - \bar{m}^{\mu}m^{\nu}$. Then now one can obtain the 
contravariant components of
the metric and the null tetrad vectors from the covariant components of the 
metric given in eq.(3) as
\begin{eqnarray}
&g^{00}&=0, ~~~g^{11}=-\lambda^2(r), ~~~g^{10}=1, \\
&g^{22}&=-{1\over r^2}, ~~~g^{33}=-{1\over r^2 \sin^2 \theta} \nonumber
\end{eqnarray}  
and
\begin{eqnarray}
&l^{\mu}& = \delta^{\mu}_{1}, ~~~n^{\mu} = \delta^{\mu}_{0} - {1\over 2}
\lambda^2(r)\delta^{\mu}_{1}, \nonumber \\
&m^{\mu}& = {1\over \sqrt{2}r}(\delta^{\mu}_{2} + {i\over \sin \theta}
\delta^{\mu}_{3}), \\
&\bar{m}^{\mu}& = {1\over \sqrt{2}r}(\delta^{\mu}_{2} - {i\over \sin \theta}
\delta^{\mu}_{3}) \nonumber
\end{eqnarray}
respectively. Now the radial coordinate $r$ is allowed to take complex values
and the tetrad is rewritten in the form
\begin{eqnarray}
&l^{\mu}& = \delta^{\mu}_{1}, ~~~n^{\mu} = \delta^{\mu}_{0} - {1\over 2}
(-M+{r\bar{r}\over l^2})\delta^{\mu}_{1}, \nonumber \\
&m^{\mu}& = {1\over \sqrt{2}\bar{r}}(\delta^{\mu}_{2} + {i\over \sin \theta}
\delta^{\mu}_{3}), \\
&\bar{m}^{\mu}& = {1\over \sqrt{2}r}(\delta^{\mu}_{2} - {i\over \sin \theta}
\delta^{\mu}_{3}) \nonumber
\end{eqnarray}
with $\bar{r}$ being the complex conjugate of $r$ (note that part of the 
algorithm
is to keep $l^{\mu}$ and $n^{\mu}$ real and $m^{\mu}$ and $\bar{m}^{\mu}$ the
complex conjugate of each other). We now formally perform the ``complex 
coordinate transformation"
\begin{eqnarray}
r' &=& r + i a \cos \theta,  ~~~\theta' = \theta, \\
u' &=& u - i a \cos \theta,  ~~~\phi' = \phi \nonumber
\end{eqnarray}
on tetrad vectors $l^{\mu}$, $n^{\mu}$ and $m^{\mu}$ ($\bar{m}'^{\mu}$ is, 
as stated, defined as the complex conjugate of ${m}'^{\mu}$). 
If one now allows $r'$ and $u'$ to be real, we obtain the following tetrad
\begin{eqnarray}
&l'^{\mu}& = \delta^{\mu}_{1}, \\
&n'^{\mu}& = \delta^{\mu}_{0} - {1\over 2}
[-M+{(r'^2+a^2\cos^2 \theta) \over l^2}]\delta^{\mu}_{1}, \nonumber \\
&m'^{\mu}& = {1\over \sqrt{2}(r'+ia\cos \theta)}[ia\sin \theta 
(\delta^{\mu}_{0} - \delta^{\mu}_{1}) +
\delta^{\mu}_{2} + {i\over \sin \theta}\delta^{\mu}_{3}] \nonumber
\end{eqnarray}
from which one can readily read off the contravariant components of the 
metric and 
then obtain the covariant components by inversion as (henceforth we drop the
``prime")
\begin{eqnarray}
ds^2_{4} &=& (-M+{\Sigma \over l^2})du^2 + 2a\sin^2 \theta [1-(-M+{\Sigma 
\over l^2})] du d\phi + 2 du dr \nonumber \\
&-& 2a\sin^2 \theta dr d\phi - \Sigma d\theta^2 -
[r^2+a^2+a^2\sin^2 \theta \{1-(-M+{\Sigma \over l^2})\}]\sin^2 \theta d\phi^2 
\end{eqnarray}
where $\Sigma \equiv (r^2+a^2\cos^2 \theta)$.
Now at this stage, considering that the geometry
in 3-dim. can be thought of as the $\theta = \pi/2$-slice of the full, 4-dim. 
one,
we simply set $\theta = \pi/2$ in the metric above to arrive at the rotating
black hole metric in 3-dim. given by
\begin{eqnarray}
ds^2 = (-M+{r^2\over l^2})(du - ad\phi)^2 
+ 2(du - ad\phi)(dr + ad\phi) - r^2 d\phi^2.
\end{eqnarray} 
Also note that the metric above we ``derived" via Newman's complex coordinate
transformation method is given in terms of Kerr-type coordinates [2,4]
$(u, r, \phi)$
which can be thought of as the generalization of the retarded null coordinates.
Thus one might want to further transform it into the one written in
Boyer-Lindquist-type coordinates [5]
$(t, r, \hat{\phi})$ that can be viewed as the generalization of the 
Schwarzschild coordinates. This can be achieved via the transformation
\begin{eqnarray}
dt = du + {(r^2+a^2)\over \Delta}dr, 
~~~d\hat{\phi} = d\phi + {a\over \Delta}dr 
\end{eqnarray}
where $\Delta \equiv r^2(-M+r^2/ l^2)+a^2$.
Finally, the rotating AdS$_{3}$ black hole solution  given in Boyer-Lindquist-
type coordinates is given by (henceforth we drop ``hat" on $\phi$ coordinate)
\begin{eqnarray}
ds^2 &=& (-M+{r^2\over l^2})dt^2 + 2a[1-(-M+{r^2\over l^2})]dt d\phi \\
&-& [r^2+a^2+a^2\{1-(-M+{r^2\over l^2})\}]d\phi^2 - {r^2\over \Delta}dr^2. \nonumber
\end{eqnarray}
And it is straightforward to check that the ``derived''
rotating black hole solution given in eq.(12) does satisfy the AdS$_{3}$
Einstein
equation, $R_{\mu \nu} - {1\over 2}g_{\mu \nu}R  - l^{-2}g_{\mu \nu} = 0$.\\
It is a little puzzling, however, that the derived rotating black hole 
solution of ours in eq.(12) above does not appear exactly the same as the one 
originally constructed by BTZ.
Nonetheless, both our solution above and the one originally
obtained by BTZ (written in our notation convention in which $a = J/2$ with 
$J$ appearing in original BTZ's work [1])
\begin{eqnarray}
ds^2 = (-M+{r^2\over l^2})d\tilde{t}^2 + 2ad\tilde{t}d\phi - r^2d\phi^2 - 
{r^2\over \Delta}dr^2
\end{eqnarray}
correctly reduce, in the vanishing angular momentum limit $a\rightarrow 0$, to
the static, spherically-symmetric black hole solution which was the starting
point of our solution construction. 
Therefore, presumably the two rotating black hole solutions must be related to
each other by a coordinate transformation.
Indeed, one can check straightforwardly that the two are related by the 
transformation of the time coordinate alone
\begin{eqnarray}
\tilde{t} = t - a\phi.
\end{eqnarray} 
Therefore the two metrics given in eqs.(12) and (13) represent one and the same
AdS$_{3}$ black hole solution modulo gauge transformation.
\\
{\bf III. Boyer-Lindquist-type coordinates versus BTZ coordinates}
\\
It is interesting to note that the rotating BTZ black hole solution given in the
Boyer-Lindquist-type time coordinate in eq.(12) resembles the structure
of Kerr solution in 4-dim. more closely than that given in the BTZ time
coordinate in eq.(13).
Therefore, it seems that now we are left with the question ; 
between $t$ and $\tilde{t}$,
which one is the usual Killing time coordinate ?
It appears that it is $\tilde{t}$, the BTZ time coordinate that is
the usual Killing time coordinate, {\it not} $t$, the Boyer-Lindquist-type 
time coordinate. And this conclusion is based on the following observation.
Note that both in Boyer-Lindquist-type coordinates $(t, ~r, ~\phi)$ and in
BTZ coordinates $(\tilde{t}, ~r, ~\phi)$, generally $\mid g_{t\phi} \mid$
$(\mid g_{\tilde{t}\phi} \mid)$ represents {\it angular momentum per unit 
mass} as observed by an accelerating observer at some fixed $r$ and $\phi$. 
(We stress here that the quantity we are about to introduce is the
angular momentum {\it per unit mass} at some point $r$. 
Generally, it should be
distinguished from the total angular momentum of the (4-dim.)
spacetime measured in the asymptotic region, 
$\tilde{J} = (16\pi )^{-1}\int_{S}\epsilon_{\mu\nu\alpha\beta}
\nabla^{\alpha}\psi^{\beta}$ in the notation convention of ref.[6]
with $S$ being a large sphere in the asymptotic region and
$\psi^{\mu} = (\partial/\partial \phi)^{\mu}$ being the rotational
Killing field. Thus in these definitions, angular momentum {\it per
unit mass} may change under a coordinate transformation although
the total angular momentum $\tilde{J}$ remains coordinate independent.)
To see this quickly, one only needs to write the rotating black hole metric
in ADM's (2+1)-split form
\begin{eqnarray}
ds^2 = N^2(r)dt^2 - f^{-2}(r)dr^2 - R^2(r)[N^{\phi}(r)dt + d\phi]^2.
\end{eqnarray}
Then in Boyer-Lindquist-type time coordinate $t$, metric components
correspond to
\begin{eqnarray}
f^{2}(r) &=& (-M + {r^2 \over l^2} + {a^2 \over r^2}), \nonumber \\
R^2(r) &=& [r^2 + a^2 +a^2 \{1 - (-M + {r^2 \over l^2})\}], \\
N^2(r) &=& (-M + {r^2 \over l^2}) + R^{-2}(r) a^2 \{1 - (-M + {r^2 \over
l^2})\}^2, \nonumber \\
N^{\phi}(r) &=& -R^{-2}(r) a \{1 - (-M + {r^2 \over l^2})\} \nonumber
\end{eqnarray}
whereas in BTZ time coordinate $\tilde{t}$, they correspond to
\begin{eqnarray}
N^2(r) &=& f^{2}(r) = (-M + {r^2 \over l^2} + {a^2 \over r^2}), \nonumber\\
N^{\phi}(r) &=& -{a \over r^2}, ~~~~R^2(r) = r^2. \nonumber
\end{eqnarray}
Now it is apparent in this ADM's space-plus-time split form of the
metric given in eq.(15) that the shift function $N^{\phi}(r)$ corresponds
to the angular velocity $\mid N^{\phi}(r) \mid = \Omega(r)$ and 
obviously $g_{\phi \phi} = R^2(r)$ represents (radius associated with the
proper circumference)$^2$. That is, 
$\int_{0}^{2\pi}(g_{\phi \phi})^{1/2} d\phi = 2\pi R(r)$ is the
proper circumference of a circular orbit around the axis of rotation. 
Therefore the 
quantity $R^2(r) \mid N^{\phi}(r) \mid = J$ can be identified with
the {\it angular momentum per unit mass at some point (with fixed $r$ 
and $\phi$) from the axis of rotation.} \\
Thus from $-R^2(r)N^{\phi}(r) = g_{t\phi}$, the angular momentum 
per unit mass in
Boyer-Lindquist-type time coordinate and in BTZ time coordinate are
given respectively by
\begin{eqnarray}
J &=& \mid g_{t\phi} \mid = a [1 - (-M + {r^2 \over l^2})], \nonumber \\
J^{BTZ} &=& \mid g_{\tilde{t} \phi} \mid = a. 
\end{eqnarray}
Now from eq.(17), it is manifest that the angular 
momentum per unit mass given in BTZ time coordinate $\tilde{t}$ is
finite and constant all over the hypersurface whereas that given
in Boyer-Lindquist-type time coordinate $t$ grows indefinitely
as $r \rightarrow \infty$.
This can be attributed to the fact that the coordinate $\tilde{t}$
BTZ used is defined asymptotically using the asymptotic symmetries
and hence approaches the AdS time at spatial infinity [1].
Therefore we can conclude that it is the BTZ time coordinate
$\tilde{t}$ which is the usual Killing time. This is certainly
in contrast to what happens in the familiar Kerr black hole
geometry in 4-dim. where the usual Boyer-Lindquist time coordinate
is the Killing time coordinate. And it seems that this discrepancy 
comes from the fact that the 3-dim. BTZ black hole is not
asymptotically flat but asymptotically anti-de Sitter whereas
the 4-dim. Kerr black hole is asymptotically flat.
The next question one might want to ask and answer could then be ; what is
the relative physical meaning of these two time coordinates 
$t$ and $\tilde{t}$ ?
To get a quick answer to this question, we go back and look at the
coordinate transformation law given in eq.(14) relating the two
time coordinates $t$ and $\tilde{t}$. Namely, taking the dual of
the transformation law
$\delta \tilde{t} = \delta t - a \delta \phi$, we get
\begin{eqnarray}
({\partial \over \partial \tilde{t}})^{\mu} &=&
({\partial \over \partial t})^{\mu} - {1\over a}
({\partial \over \partial \phi})^{\mu}  \\
{\rm or} \nonumber \\
\tilde{\xi}^{\mu} &=& \xi^{\mu} - {1\over a}\psi^{\mu} \nonumber
\end{eqnarray}
where $\xi^{\mu} = (\partial /\partial t)^{\mu}$ and
$\psi^{\mu} = (\partial /\partial \phi)^{\mu}$ denote Killing 
fields corresponding to the time translational and the rotational
isometries of the spinning black hole spacetime respectively and
$\tilde{\xi}^{\mu} = (\partial /\partial \tilde{t})^{\mu}$ denotes
the Killing field associated with the isometry of the hole's metric
under the BTZ time translation. Now this expression for the BTZ
time translational Killing field $\tilde{\xi}^{\mu}$ implies that
in BTZ time $\tilde{t}$, the time translational generator is
given by the linear combination of the Boyer-Lindquist-type time 
translational generator and the rotational generator. In plain
English, this means that in BTZ time coordinate, the action of
time translation consists of the action of Boyer-Lindquist-type time
translation and the action of rotation in opposite direction to
$a$, i.e., to the rotation direction of the hole. Thus the BTZ
time coordinate $\tilde{t}$ can be interpreted as the coordinate,
say, of a frame which rotates around the axis of the spinning
BTZ black hole in opposite direction to that of the hole outside
its static limit. \\
Finally, we come to another peculiar point worthy of note.
The rotating BTZ black hole metric, when represented in
Boyer-Lindquist-type coordinates as given in eq.(12), exhibits
the following exotic behavior. Notice that in the Boyer-Lindquist-
type coordinates, the rotational Killing field 
$\psi^{\mu} = (\partial /\partial \phi)^{\mu}$ fails to remain
strictly spacelike while it does remain everywhere spacelike
in BTZ coordinates. That is, from
\begin{eqnarray}
\psi^{\mu}\psi_{\mu} = g_{\phi \phi} &=& r^2(1-a^2/l^2) +
a^2(M+2) \nonumber \\
&=& r^2(1-|\Lambda|a^2) + a^2(M+2) 
\end{eqnarray}
(where we used $l^{-2} = -\Lambda = |\Lambda| > 0$), we can
realize that ; \\
(i) for $a < 1/\sqrt{|\Lambda|}$ (i.e., slowly spinning hole),
$g_{\phi \phi}>0$, namely, $\psi^{\mu}$ is everywhere spacelike
and hence the ``proper circumference'' of a circular orbit 
around the axis of rotation, 
$\int_{0}^{2\pi}(g_{\phi \phi})^{1/2} d\phi = 
2\pi (g_{\phi \phi})^{1/2}$,
is positive-definite and increases with $r$, and \\
(ii) for $a = 1/\sqrt{|\Lambda|}$, $g_{\phi \phi} = (M+2)a^2 
= const.$, namely, $\psi^{\mu}$ is again everywhere spacelike
but the proper circumference $2\pi (g_{\phi \phi})^{1/2}$ remains
constant at all distances from the axis of rotation, and lastly \\
(iii) for $a > 1/\sqrt{|\Lambda|}$ (i.e., rapidly spinning hole),
$g_{\phi \phi} = - (|\Lambda|a^2 - 1)r^2 + (M+2)a^2$ could become
negative-definite for
$r > a\sqrt{{M+2 \over |\Lambda|a^2 - 1}}$ or for
$r < - a\sqrt{{M+2 \over |\Lambda|a^2 - 1}}$ ($r$ may be extended
to ``negative'' values as we shall see in the next section).
This implies that in these ``far'' regions, the $\phi$-coordinate
could become timelike which, in turn, leads to the occurrence of
the closed timelike curves signaling the possible violation of
causality. This is a reminiscence of what happens in the case
of Kerr black hole spacetime. Namely, as was first pointed out
by Carter [5], the closed timelike curves can occur (i.e.,
$\psi^{\mu}$ can go timelike) as one approaches from negative-r
region toward the ring singularity on the equatorial plane of
Kerr spacetime. And this could be another evidence that it is
the BTZ time coordinate, not that of Boyer-Lindquist-type, which
bears better physical relevance. 
\\
{\bf IV. Kerr-Schild-type coordinates and the maximal extension
 of global structure}
\\
Thus far, we have attempted the analysis of the AdS$_3$ black hole
solution by BTZ in a fashion which is parallel with that of Kerr
black hole in 4-dim. Namely, starting from the nonrotating BTZ
black hole solution, we derived the rotating solution via Newman's
complex coordinate transformation method. Then, similarly to what
happens in the case of Kerr black hole in 4-dim., we obtained, as
a result of this Newman's algorithm, the rotating
BTZ black hole solution first in Kerr-type coordinates $(u, r, \phi)$
given in eq.(10). By performing the transformation in eq.(11), next 
we represented the black hole metric in Boyer-Lindquist-type
coordinates $(t, r, \hat{\phi})$ given in eq.(12). Therefore in
order to complete our study of rotating BTZ black hole solution
in parallel with that of Kerr solution, we are naturally led to the
remaining final step. That is, we look for further transformation
to Kerr-Schild-type coordinates just as we did in the study of Kerr
solution in 4-dim. Thus at this point, it seems appropriate to
briefly review the physical roles played by these three alternative
coordinate systems in which the rotating Kerr metric was
represented. Historically, the Kerr solution was originally found
in Kerr coordinates [2] and then the Kerr-Schild coordinates [3]
was introduced and finally the Boyer-Lindquist coordinates was
discovered [5]. Firstly, the Kerr coordinates $(u, r, \theta, \phi)$
can be thought of as the generalization of Eddington-Finkelstein
null coordinates and hence is free of coordinate singularities.
Next, the Boyer-Lindquist coordinates $(t, r, \theta, \hat{\phi})$ can 
be viewed as the generalization of (accelerated) Schwarzschild coordinates.
In this system, it is straightforward to see how the charged Kerr metric 
reduces to the
familiar Schwarzschild $(a = e = 0)$ or Reissner-Nordstrom $(a = 0)$
metrics (where $a$ and $e$ denote the angular momentum per unit mass
and the electric charge respectively). It is also clear that the
spacetime is asymptotically-flat in the limit of large positive or
negative values of $r$. Lastly, the Kerr-Schild coordinates
$(\tilde{t}, x, y, z)$ is a quasi-Cartesian coordinates and the well-known
``ring'' structure of curvature singularity can only be uncovered in
this coordinate system. Besides, the true maximal extension by
analytically continuing the $r$-coordinate from positive to negative
values can be performed in this system. \\
The transformation law from Kerr to Boyer-Lindquist coordinates
for Kerr solution is the same as that given in eq.(11) except that
now $\Delta = r^2 - 2Mr + a^2$. Next, the transformation law from 
Boyer-Lindquist to Kerr-Schild coordinates is given by
\begin{eqnarray}
x + iy &=& (r + ia)\sin \theta e^{i\tilde{\phi}}, \nonumber \\
z &=& r \cos \theta, ~~~\tilde{t} = v - r
\end{eqnarray}
where
\begin{eqnarray}
\tilde{\phi} = \hat{\phi} + \int dr {a\over \Delta},
~~~v = t + \int dr {{r^2+a^2}\over \Delta} \nonumber
\end{eqnarray}
and $r$ is determined implicitly by
$r^4 - (x^2 + y^2 + z^2 - a^2)r^2 - a^2 z^2 = 0$.
Then the Kerr metric given in this Kerr-Schild coordinates reads
\begin{eqnarray}
ds^2 &=& (-d\tilde{t}^2 + dx^2 + dy^2 + dz^2) \\
&+& {{r^2(r^2+a^2-\Delta)}\over {r^4+a^2 z^2}}
\left[{{r(xdx + ydy) - a(xdy - ydx)}\over {r^2 + a^2}} 
+ {zdz \over r}+ d\tilde{t} \right]^2. \nonumber
\end{eqnarray} 
Now we come back to the representation of the rotating BTZ black
hole solution in Kerr-Schild-type coordinates. When we derived the
rotating BTZ black hole solution from its nonrotating counterpart
via Newman's algorithm, our philosophy was that the 3-dim.
spacetime of interest can be thought of as, say, the $\theta =\pi/2$-
slice (i.e., the equatorial plane) of the 4-dim. one. Thus in
accordance with this philosophy, we take the coordinate transformation
law from Boyer-Lindquist-type $(t, r, \hat{\phi})$ to
Kerr-Schild-type $(\tilde{t}, x, y)$ coordinates to be the special
case of that for Kerr solution in 4-dim. given above when
$\theta =\pi/2$. Then
\begin{eqnarray}
x + iy &=& (r + ia)e^{i\tilde{\phi}}, \nonumber \\
\tilde{t} &=& v - r
\end{eqnarray}
where
\begin{eqnarray}
\tilde{\phi} = \hat{\phi} + \int dr {a\over \Delta},
~~~v = t + \int dr {{r^2+a^2}\over \Delta} \nonumber
\end{eqnarray}
with $\Delta = r^2(-M+r^2/l^2)+a^2$ now
and $r$ is determined implicitly by
$x^2 + y^2 = r^2 + a^2$.
Then the rotating BTZ black hole metric given in this Kerr-Schild-type 
coordinates reads
\begin{eqnarray}
ds^2 &=& (-d\tilde{t}^2 + dx^2 + dy^2) \\
&+& {{(r^2+a^2-\Delta)}\over {r^2}}
\left[{{r(xdx + ydy) - a(xdy - ydx)}\over {r^2 + a^2}}
+ d\tilde{t} \right]^2. \nonumber
\end{eqnarray}
Note first that the relation
\begin{eqnarray}
r^2 = x^2 + y^2 - a^2
\end{eqnarray}
implies that each set of values $(x, y)$ corresponds to two different
points distinguished by positive and negative values of $r$. This
indicates that $r$ can take negative values as well as positive
values. In fact the analytic continuation of $r$-coordinate from
positive to negative values through $r = 0$ is permissible as the
rotating BTZ black hole metric remains regular at $r = 0$. 
It is also interesting to note that this form of the rotating BTZ black
hole metric written in Kerr-Schild-type coordinates corresponds
precisely to that of the Kerr black hole metric in 4-dim. again
written in Kerr-Schild coordinates if one sets
$z = r\cos \theta = 0$. This point seems to indicate that our
philosophy, i.e., viewing the 3-dim. black hole spacetime as the
$\theta = \pi/2$-slice (equatorial plane) of a 4-dim. spacetime was 
not so naive, after all.  Besides, it further supports our earlier
statement that it is the Boyer-Lindquist-type coordinates we discovered
in the present work, not the BTZ time coordinate, in which the form
of the spinning BTZ black hole metric resembles more closely that of
the Kerr black hole metric in 4-dim. 
Lastly, we now turn to the {\it truely} maximal analytic extension
of the global structure of spinning BTZ black hole spacetime that
can be completed only when one works in this Kerr-Schild-type
quasi-Cartesian coordinates. Namely, when the spinning BTZ black hole
metric is written in this Kerr-Schild-type coordinates, it becomes
natural to consider the analytic continuation of $r$-coordinate from
positive to negative values through $r = 0$ in a similar manner to
the case of Kerr black hole in 4-dim. Even after this ``truely''
maximal analytic extension (involving the extension to negative-$r$),
however, the Carter-Penrose conformal diagram of the spinnig BTZ black hole
spacetime remains almost the same except that the spatial infinity
in the negative-$r$ side, $r = -\infty$ now should be incorporated.
Just like $r = +\infty$, $r = -\infty$ is also represented by a timelike
line and hence the slightly modified conformal diagram showing this
new ingredient is given in Fig.1. 
\\
{\bf V. Discussions}
\\
To summarize, here in this work we explored the structure of rotating
BTZ spacetime in parallel with that of Kerr spacetime in 4-dim.
And such a study was initiated from our attempt to apply Newman's method
of generating a spinning black hole solution from a static black hole
solution in 4-dim. Einstein theory to the 3-dim. situation.
More concretely, we employed the algorithm ; dimensional continuation
$\rightarrow $ Newman's derivation method $\rightarrow $ dimensional
reduction by setting $\theta = \pi/2$ to successfully obtain a legitimate
rotating AdS$_{3}$ black hole solution.
And as a consequence of this study, we discovered that there are two
alternative time coordinates in describing
the rotating AdS$_{3}$ black hole solution one can select from
to investigate its various physical contents.
Thus let us elaborate on the complementary roles played by the 
two alternative time coordinates. 
First, note that the two time coordinates, that of  Boyer-Lindquist-type,
$t$ and that of BTZ, $\tilde{t}$ coincide for nonrotating case
$(a = 0)$ and become different only for rotating case $(a \neq 0)$
as one can see in their relation, $\tilde{t} = t - a\phi$.
Next, the Boyer-Lindquist-type time coordinate
$t$ is the ``familiar'' time coordinate in which the rotating BTZ
black hole spacetime metric closely resembles that of the Kerr
(or Kerr-de Sitter [7,8]) spacetime metric in 4-dim. Thus for direct and 
parallel comparison 
between the structure of 3-dim. rotating BTZ black hole and that of 4-dim.
Kerr (or Kerr-de Sitter) black hole, it seems appropriate to work
in the Boyer-Lindquist-type time coordinate.
The BTZ time coordinate $\tilde{t}$, on the other hand, may look
``unusual''
in that it can be identified with the time coordinate of a non-static
observer who rotates in opposite direction to that of the spinning hole.
It is, however, this BTZ time coordinate, not the familiar
Boyer-Lindquist-type time coordinate, which is the right Killing
time coordinate in terms of which the total mass and particularly
the  angular momentum can be well-defined in the asymptotic
region of this asymptotically anti-de Sitter spacetime.
Also it has been well-studied in the literature [1] that this BTZ time 
coordinate
is particularly advantageous in exploring the global structure of the
rotating BTZ black hole since it allows one to transform to Kruskal-type 
coordinates and eventually allows one to  draw the Carter-Penrose 
conformal diagram much more easily than the case when one employs 
the usual Boyer-Lindquist-type time coordinate. 
Therefore when investigating various physical contents of the spinning
BTZ black hole, the two time coordinates $t$ and $\tilde{t}$ appear
to play mutually complementary roles. Lastly, the discovery of the
subsequent transformation from Boyer-Lindquist-type to Kerr-Schild-type
quasi-Cartesian coordinates allowed, among other things, us to 
further uncover the hidden rich global structure of the rotating
BTZ black hole spacetime. Namely, in the Kerr-Schild-type coordinates,
it became natural to maximally extend the global structure by
analytically continuing to ``antigravity universe'' region (i.e.,
negative $r$-region) just as we did in the case of Kerr black hole
spacetime in 4-dim.

\vspace{2cm}

{\bf Acknowledgement}
\\
The author would like to thank Dr. Hee-Won Lee for generating the
Carter-Penrose conformal diagram of maximally-extended rotating
BTZ spacetime.

\vspace{2cm}

{\bf References}

\begin{description}

\item {[1]} M. ~Ba$\tilde{\rm{n}}$ados, C. ~Teitelboim, and J. ~Zanelli, 
Phys. Rev. 
Lett. {\bf 69}, 1849 (1992) ;  M. ~Ba$\tilde{\rm{n}}$ados, M. ~Henneaux, 
C. ~Teitelboim, and J. ~Zanelli, Phys. Rev. {\bf D48}, 1506 (1993).
\item {[2]} R. P. Kerr, Phys. Rev. Lett. {\bf 11}, 237 (1963).
\item {[3]} R. P. Kerr and A. Schild, Am. Math. Soc. Symposium,
New York, 1964.
\item {[4]} E. T. ~Newman and A. I. ~Janis, J. Math. Phys. {\bf 6}, 915 
(1965) ;
E. T. ~Newman, E. ~Couch, R. ~Chinnapared, A. ~Exton, A. ~Prakash and
R. ~Torrence, J. Math. Phys. {\bf 6}, 918 (1965).
\item {[5]} R. H. ~Boyer and R. W. ~Lindquist, J. Math. Phys. {\bf 8}, 265 
(1967) ; B. ~Carter, Phys. Rev. {\bf 174}, 1559 (1968).
\item {[6]} R. M. ~Wald, {\it General Relativity} (Univ. of Chicago Press, 
Chicago, 1984).
\item {[7]} B. ~Carter, Commun. Math. Phys. {\bf 17}, 233 (1970) ;
B. ~Carter, in {\it Les Astre Occlus} (Gordon and Breach, New York, 1973).
\item {[8]} G. W. ~Gibbons and S. W. ~Hawking, Phys. Rev. {\bf D15}, 2738 
(1977).

\end{description}

\newpage

{\bf Figure Caption}
\\
\\
\\
\\
FIG.1 Carter-Penrose conformal diagram of {\it truely} maximally-extended
      spinning BTZ black hole spacetime. Note that it is almost the same
      as that before the analytic continuation to negative-$r$  except for 
      the emergence of negative-$r$ region with a new timelike infinity
      at $r = -\infty$.

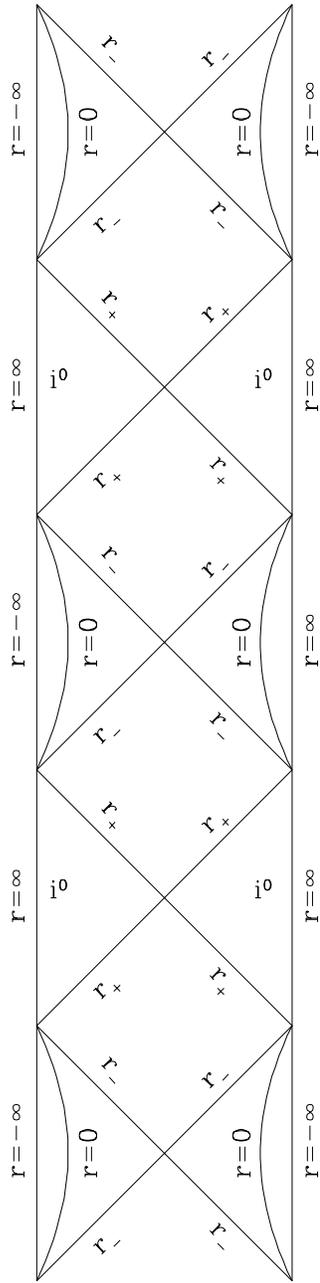
\end{document}